# Interfacial Hole Accumulation Enables Selective Free-Surface Formation in Ge-on-Si

Youcef A. Bioud[1, a)], Meriem Bouchilaoun[2], Waldemar Schreiber[1], Redouane Amrar[2,3], Gilles Patriarche[4], Ali Soltani[2,3], and Stefan Janz[1, b)]

*[1]Division Photovoltaics, Fraunhofer Institute for Solar Energy Systems (ISE), Heidenhofstraße 2, 79110 Freiburg, Germany.*
*[2]Institut Interdisciplinaire d'Innovation Technologique (3IT), Université de Sherbrooke, 3000 Boulevard de l'Université, Sherbrooke, QC J1K 0A5, Canada.*
*[3]Laboratoire Nanotechnologies Nanosystèmes (LN2) – IRL3463, CNRS, Université de Sherbrooke, INSA Lyon, École Centrale de Lyon, Université Grenoble Alpes, Sherbrooke, QC J1K 0A5, Canada.*
*[4]Université Paris-Saclay, CNRS, Centre de Nanosciences et de Nanotechnologies, 91120 Palaiseau, France.*

*[a)]Corresponding author: y.bioud@usherbrooke.ca*
*[b)]stefan.janz@ise.fraunhofer.de*

**Abstract.** Electrochemically formed two-dimensional free surfaces (2DFS) decouple Ge epilayers from the misfit-dislocation network at the Ge/Si interface, but the origin of their selectivity has remained unclear. We combine finite-volume current-distribution modelling, including charge-transfer kinetics, with a one-dimensional Poisson analysis of the p-Ge/$p^+$-Si heterojunction. With realistic conductivities, an ohmic description would drive 77% of the anodic current through Ge, whereas hole-limited charge transfer reduces this fraction to 0.07%, independently of the electrolyte conductivity. The valence-band offset accumulates holes on the Ge side up to ~$10^{19}$ $cm^{-3}$ within a self-limited nanometric layer, which localizes dissolution at the interface. The model yields design rules for the doping contrast ($N_{Si}/N_{Ge} \gtrsim 60$) and the mesa width ($\lesssim 3.6\lambda$, with $\lambda \approx 1.7$ µm).

## INTRODUCTION

Ge-on-Si is a key platform for Si photonics and III–V integration on Si,[1,2] but the 4.2% lattice mismatch generates misfit and threading dislocations. We have shown that porous and voided Ge/Si interfaces, created by electrochemical etching and subsequent annealing, can trap and annihilate threading dislocations,[3–8] and device modelling has shown their potential for III–V multi-junction solar cells on Si;[9,10] this work builds on bipolar electrochemical etching of Ge.[11,12] This concept was recently extended to a nanometric 2DFS that decouples the epilayer from the substrate.[13] Earlier current-distribution simulations of Si electrochemical etching treated the semiconductor/electrolyte boundary as ohmic.[14] Here, we add charge-transfer kinetics and show that two effects act together: hole-limited charge transfer, set by the doping contrast, delivers the current to Si and the buried interface, and hole accumulation at the heterojunction makes the interface the most reactive site.

## MODEL

The current distribution was computed for an axisymmetric mesa unit cell [Fig. 1(a)]: a 3-µm-wide mesa made of 3 µm of p-Ge ($10^{15}$ $cm^{-3}$) on a 3-µm-high $p^+$-Si pedestal ($10^{18}$ $cm^{-3}$), with a top mask, in HF/ethanol electrolyte, solved by a finite-volume scheme refined to 0.2 nm near the interface. Bulk conductivities follow from $\sigma = qN_A\mu_p$, using the Masetti mobility model for Si,[15] which gives $\sigma_{Ge} \approx 30$ S $m^{-1}$ and $\sigma_{Si} \approx 2.4 \times 10^3$ S $m^{-1}$; the electrolyte conductivity $\sigma_{el} = 10$ S $m^{-1}$ was varied from 1 to $10^4$ S $m^{-1}$. Because anodic dissolution of p-type semiconductors consumes holes supplied at the surface,[12] each semiconductor/electrolyte boundary carries a linearized charge-transfer resistance in series with the bulk, $r_{ct} = r_{ct,ref}(10^{18}\ cm^{-3}/N_A)$, with $r_{ct,ref} = 2\ \Omega\ cm^2$. The equilibrium band diagram of the heterojunction [Fig. 1(b)] was obtained by solving the one-dimensional Poisson equation at 300 K with Fermi–Dirac statistics and a valence-band offset $\Delta E_v \approx 0.5$ eV, varied from 0.3 to 0.7 eV.

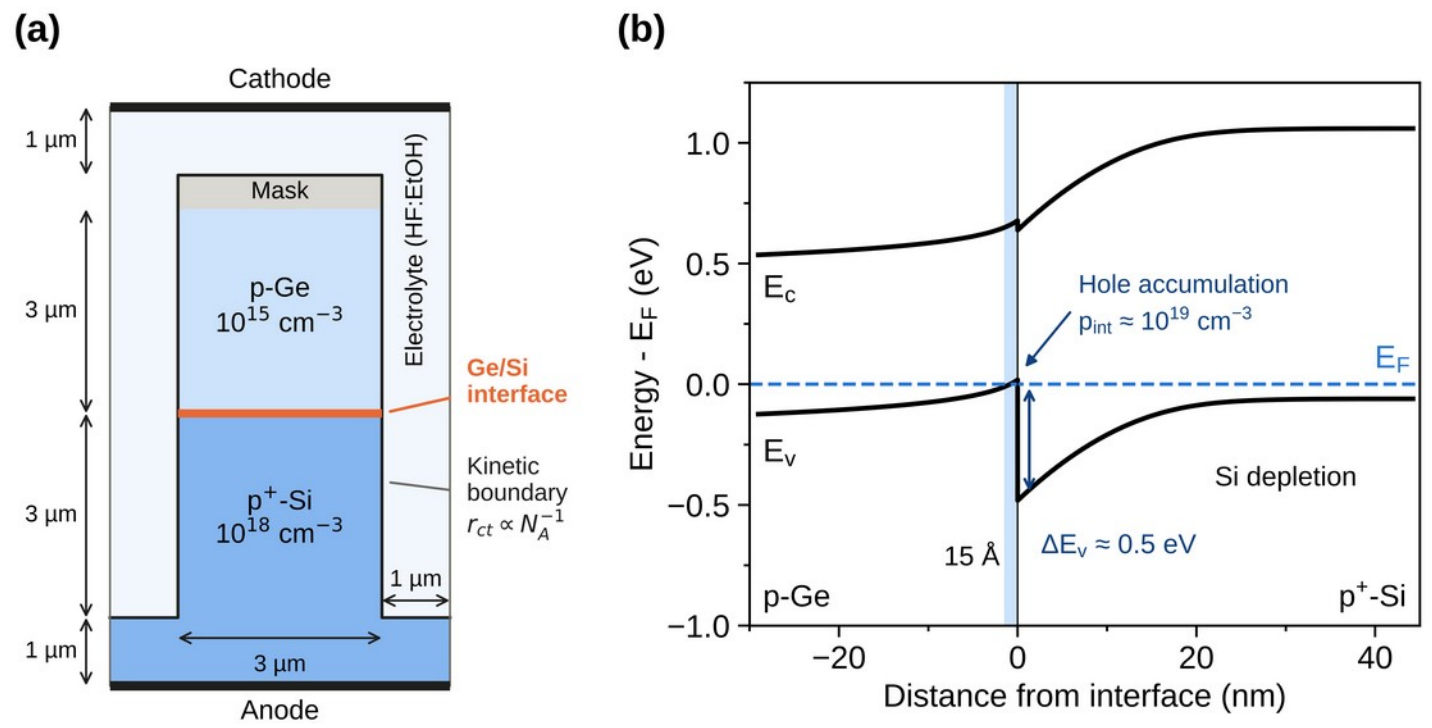

**FIGURE 1.** (a) Axisymmetric model geometry; charge-transfer kinetics with $r_{ct} \propto N_A^{-1}$ apply at all semiconductor/electrolyte boundaries. (b) Equilibrium band diagram of the p-Ge ($10^{15}$ $cm^{-3}$)/$p^+$-Si ($10^{18}$ $cm^{-3}$) heterojunction, showing hole accumulation at the interface; the shaded band marks the 15 Å 2DFS.

## RESULTS AND DISCUSSION

***Current confinement by hole-limited charge transfer.*** With an ohmic boundary and realistic conductivities, 77% of the anodic current leaves through the Ge sidewall, because $\sigma_{Ge}$ exceeds $\sigma_{el}$ [Fig. 2(a)]; Ge would be attacked, unlike the intact Ge layers observed experimentally.[13] With charge-transfer kinetics, this fraction falls to 0.07% [Fig. 2(b)]: the current bypasses the low-doped Ge and reaches the Si sidewall and the Ge/Si boundary. This result is insensitive to $\sigma_{el}$ between 1 and $10^4$ S m$^{-1}$, to a 1-nm conductive interfacial layer, and to changes of the bulk conductivities by factors of 0.25–4, because the charge-transfer resistance dominates over the bulk resistance at the micrometre scale.

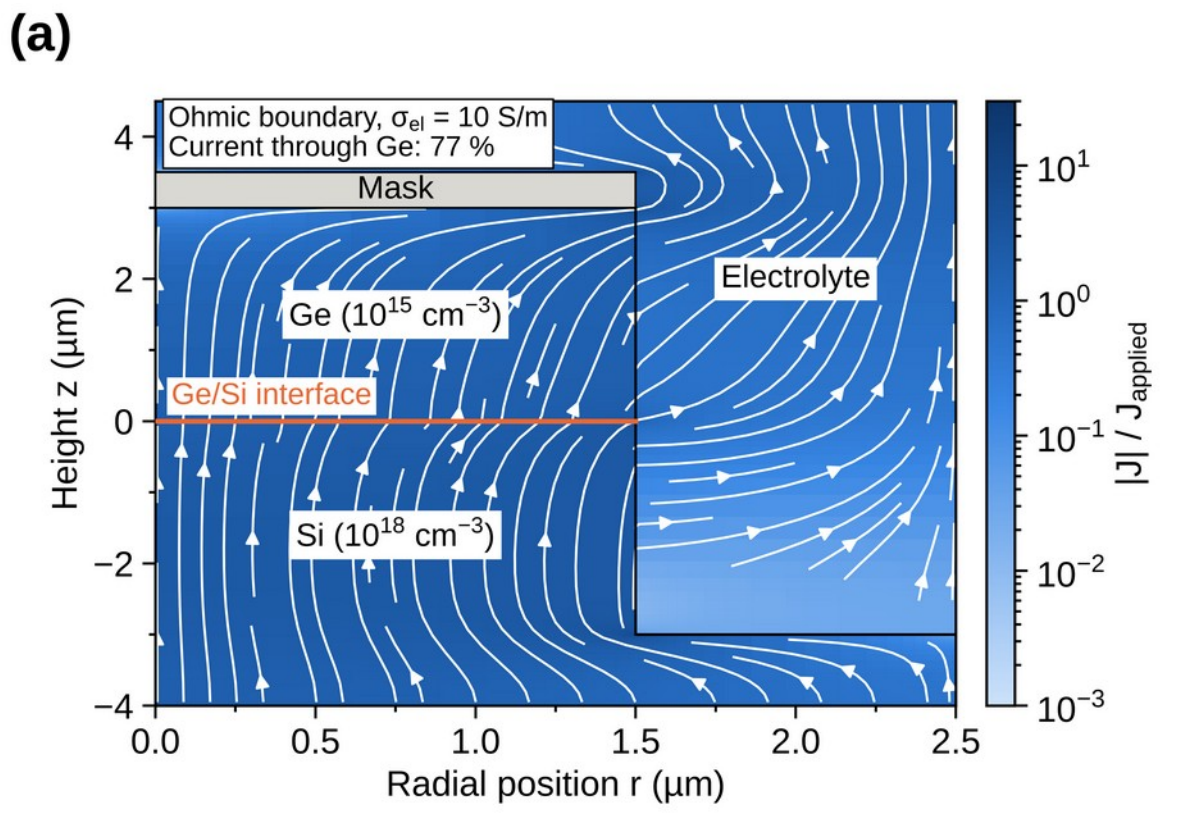

**FIGURE 2.** Simulated current-density distributions ($\sigma_{el}$ = 10 S m$^{-1}$) for (a) an ohmic boundary and (b) charge-transfer kinetics. Streamlines show the current path, and the orange line marks the Ge/Si interface.

***Interfacial hole accumulation.*** Current confinement explains why Ge is preserved, but not why dissolution follows the interface. The Poisson calculation shows that the valence-band offset transfers holes from p$^+$-Si into Ge, producing an accumulation layer with $p_{int} \approx 1.0 \times 10^{19}$ cm$^{-3}$ at the interface, about ten times the substrate concentration, and a ~20-nm depletion region on the Si side [Figs. 1(b) and 3(a)]. Because $r_{ct}$ scales as $1/p$, the interface is locally 6–15 times more reactive than the surrounding Si for $\Delta E_v$ = 0.3–0.7 eV, enough for the front to advance several times faster than Si is porosified. The hole-enriched Ge layer is ≈5 nm thick for $10^{18}$ cm$^{-3}$ and decreases to ≈1.7 nm for $10^{19}$ cm$^{-3}$ [Fig. 3(b)], with a weak dependence on Ge doping. This self-limited reactive layer is consistent in order of magnitude with the ~15 Å separation observed by STEM,[13] and predicts a thinner 2DFS for higher substrate doping.

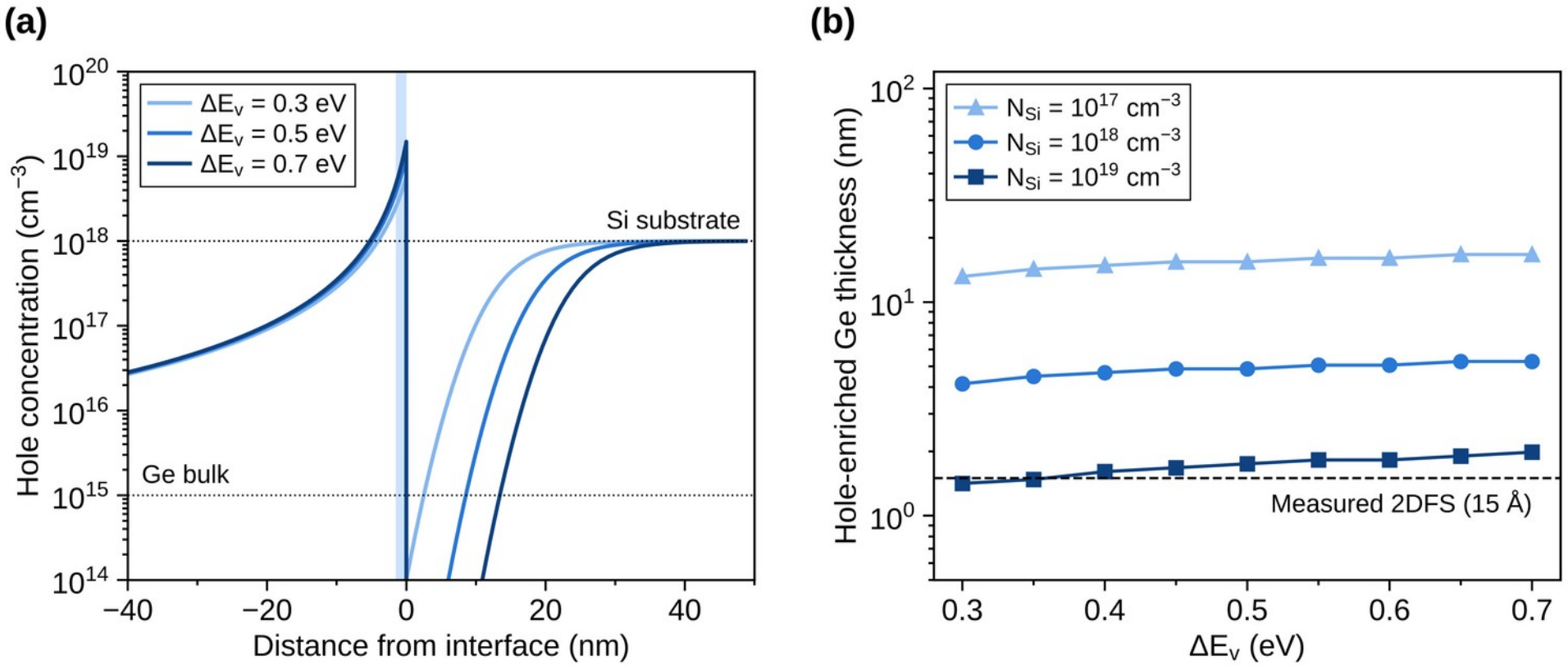

**FIGURE 3.** (a) Hole concentration across the p-Ge/p$^+$-Si heterojunction for $\Delta E_v$ = 0.3–0.7 eV; the shaded band marks the 15 Å 2DFS. (b) Thickness of the Ge layer with $p \geq N_{Si}$ versus $\Delta E_v$ for three Si doping levels; the dashed line marks the measured 2DFS thickness.

***Design rules.*** Keeping the current through Ge below 1% requires $N_{Si}/N_{Ge} \gtrsim 60$ [Fig. 4(a)]: for $10^{15}$ cm$^{-3}$ Ge, substrates of $10^{17}$, $10^{18}$, and $10^{19}$ cm$^{-3}$ give 0.7%, 0.07%, and 0.007%, respectively, and $10^{18}$ cm$^{-3}$ avoids the excessive porosification of more heavily doped Si. Once the interface opens, the current feeding the front flows through the thin electrolyte-filled gap of thickness $h$ and decays over $\lambda = (\sigma_{el} h r_{ct})^{1/2}$, i.e., $\lambda \approx 1.7$ μm for $h$ = 1.5 nm. The centre of a 3-μm mesa still receives ≈84% of the sidewall current density, compared with ≈20% for a 10-μm mesa and ≈2% for a 20-μm mesa. Requiring at least 50% at the mesa centre gives a maximum width $w_{max} \approx 3.6\lambda$ [Fig. 4(b)], i.e., 3–14 μm for $r_{ct}$ = 0.5–10 Ω cm$^2$, consistent with the 3 × 3 μm$^2$ mesas used experimentally.

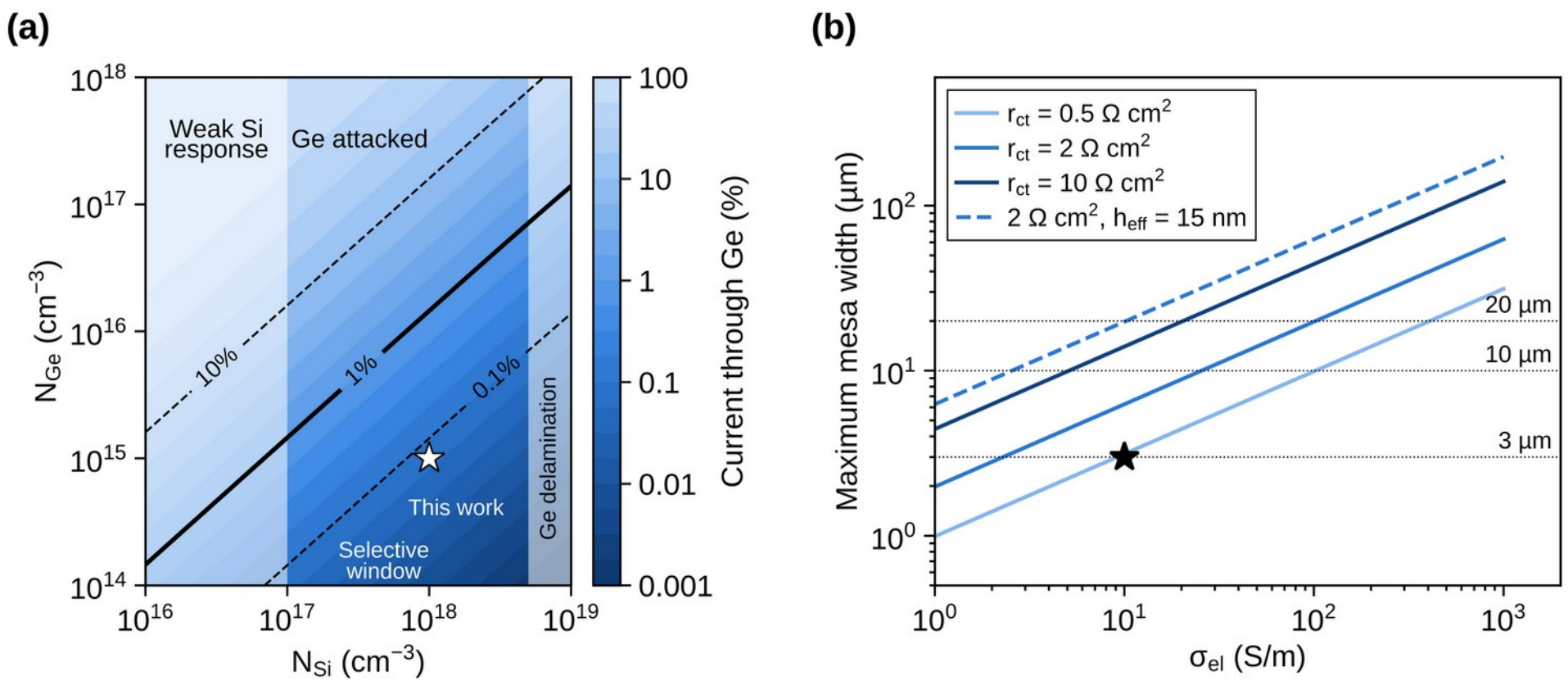

**FIGURE 4.** (a) Simulated current through Ge versus Si and Ge doping; contours mark 0.1%, 1%, and 10%, shaded bands indicate the weak Si response and Ge delamination, and the star marks the operating point. (b) Maximum mesa width for complete decoupling versus electrolyte conductivity for different charge-transfer resistances and effective gap thicknesses; the star marks 3-µm mesas at $\sigma_{el} = 10$ S m$^{-1}$.

## CONCLUSION

Selective free-surface formation at the Ge/Si interface results from two complementary effects. Hole-limited charge transfer, set by the doping contrast, confines the anodization current to Si and the heterointerface and reduces Ge attack to 0.07%, independently of the electrolyte conductivity. Interfacial hole accumulation, driven by the valence-band offset, then makes the heterojunction the most reactive site within a self-limited, nanometre-thick layer, so that dissolution propagates laterally along the interface. The resulting design rules on doping contrast and mesa width can guide buried free-surface engineering in other lattice-mismatched systems.

## ACKNOWLEDGMENTS

Y. A. Bioud gratefully acknowledges support from the German Academic Exchange Service (DAAD).